\documentstyle[12pt,axodraw]{article}
\makeatletter
\def\vereq#1#2{\lower3pt\vbox{\baselineskip1.5pt \lineskip1.5pt
\ialign{$\m@th#1\hfill##\hfil$\crcr#2\crcr\sim\crcr}}}
\makeatother

\begin{document}

\begin{titlepage}

\begin{flushright}
UCB-PTH-01/39 \\
LBNL-48987 \\
\end{flushright}

\vskip 1.5cm

\begin{center}
{\Large \bf A Constrained Standard Model: \\
             Effects of Fayet-Iliopoulos Terms}

\vskip 1.0cm

{\bf
Riccardo Barbieri$^{a}$,
Lawrence J.~Hall$^{b,c}$,
Yasunori Nomura$^{b,c}$
}

\vskip 0.5cm

$^a$ {\it Scuola Normale Superiore and INFN, Piazza dei Cavalieri 7, 
                 I-56126 Pisa, Italy}\\
$^b$ {\it Department of Physics, University of California,
                 Berkeley, CA 94720, USA}\\
$^c$ {\it Theoretical Physics Group, Lawrence Berkeley National Laboratory,
                 Berkeley, CA 94720, USA}

\vskip 1.0cm

\abstract{
In \cite{Barbieri:2001vh} the one Higgs doublet standard model was 
obtained by an orbifold projection of a 5D supersymmetric theory 
in an essentially unique way, resulting in a prediction for the 
Higgs mass $m_H = 127 \pm 8$ GeV and for the compactification scale 
$1/R = 370 \pm 70$ GeV. The dominant one loop contribution to the 
Higgs potential was found to be finite, while the above uncertainties 
arose from quadratically divergent brane $Z$ factors and from other 
higher loop contributions.  In \cite{Ghilencea:2001bw}, a quadratically 
divergent Fayet-Iliopoulos term was found at one loop in this theory. 
We show that the resulting uncertainties in the predictions
for the Higgs boson mass and the compactification scale are small, 
about 25\% of the uncertainties quoted above, and hence do not 
affect the original predictions.  However, a tree level brane 
Fayet-Iliopoulos term could, if large enough, modify these predictions, 
especially for $1/R$.}

\end{center}
\end{titlepage}

In ref.~\cite{Barbieri:2001vh} we obtained the 1 Higgs doublet standard 
model from a 5D supersymmetric theory with both the standard model gauge 
particles and the Higgs boson in the bulk.  The Scherk-Schwarz (SS) 
mechanism is employed to remove unwanted particles and symmetries: 
the entire superpartner spectrum, as well as other particles implied 
by 5D Lorentz invariance, are raised to the compactification scale, 
$1/R$. In ref.~\cite{Barbieri:2001dm} we demonstrated that this is 
the unique such construction in 5D, up to a small deformation in the 
orbifold boundary condition. 

A relevant property of the model is that the Higgs potential is 
calculable in terms of the compactification scale 1/R, up to small 
effects from supersymmetric counterterms. The Fermi constant determines
$1/R \approx 370$ GeV, so that the Higgs mass is predicted. The
leading 1 loop diagrams for the Higgs potential are exponentially
insensitive to physics at energies above the compactification scale,
but UV sensitivities can arise through the supersymmetric
counterterms, which must therefore be studied.
In ref.~\cite{Barbieri:2001vh} the leading counterterms were found 
to be brane $Z$ factors for the top quark superfields. These are 
quadratically divergent, and affect the masses of the Kaluza-Klein (KK) 
modes of the top quarks and squarks, which enter the radiative diagrams 
for the Higgs potential. The Higgs mass and compactification scale 
therefore have a sensitivity to unknown UV physics via a quadratic 
divergence at the two loop level, introducing uncertainties in the 
predictions for the Higgs mass and $1/R$ of 1\% and 20\%, respectively.
Further uncertainties in the Higgs boson mass of about 6\% arise from 
other higher loop contributions.  In ref.~\cite{Ghilencea:2001bw} 
a quadratically divergent Fayet-Iliopoulos (FI) term was noticed, 
associated with standard hypercharge, which escaped our attention. 
This introduces a quadratic sensitivity to unknown UV physics at 
the one loop level, and its consequences are studied in this note.

\begin{figure}
\begin{center} 
\begin{picture}(250,90)(-165,150)
  \Vertex(-40,174){2}
  \Line(-39,150)(-39,174) \Line(-41,150)(-41,174) 
  \Text(-30,160)[l]{$D_Y$}
  \DashCArc(-40,203)(29,0,360){3} 
  \Text(-40,240)[b]{$h, h^c$}
\end{picture}
\caption{One-loop diagram generating the FI $D$-term.}
\label{fig:diag}
\end{center}
\end{figure}
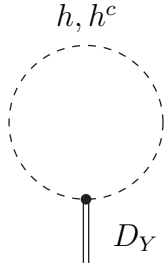

It is indeed immediate to see that the diagram of fig.~\ref{fig:diag}, 
properly calculated for the different KK components of the hypercharge 
$D$-term, gives rise to an effective Lagrangian term
\begin{equation}
  {\cal L}_{\rm eff} = 
    \frac{\xi}{\sqrt{2}} \left( \delta(y) + \delta(y-\pi R/2) \right) D_Y,
\label{eq:L-D}
\end{equation}
where
\begin{equation}
  \xi \simeq {g' \over 2} \int \frac{d^4p}{(2\pi)^4} \frac{1}{p^2} 
      \simeq {g' \over 2} \frac{\Lambda^2}{16 \pi^2},
\label{eq:xi}
\end{equation}
and
\begin{equation}
  D_Y =  \frac{D_0}{\sqrt{2}} + \sum_{n=1}^\infty D_n \cos {2ny \over R},
\label{eq:DY}
\end{equation}
with $D_0$ and $D_n$ canonically normalized in 4D.
$\Lambda$ is an ultraviolet cutoff and $g'$ is the $U(1)$ hypercharge 
coupling. In the loop of fig.~\ref{fig:diag}, only the Higgs zero-mode 
contributes to the zero-mode of $D_Y$, whereas the $2n$-th KK modes of 
$D_Y$ receive contributions from the $n$-th KK modes of the Higgs field 
and of its charge conjugate.  The matter hypermultiplets do not
contribute to a FI term on the brane due to ${\rm Tr}Y=0$ over the 
standard matter multiplets.

In the appendix, we comment on the theoretical issues 
about the generation of the FI term.

It is important to notice that ${\cal L}_{\rm eff}$ 
is perfectly compatible with the residual 
supersymmetries of the full Lagrangian after the orbifold projection. 
This Lagrangian, other than the fully supersymmetric term in 5D, 
${\cal L}_5$, must include the most general 4D Lagrangians at $y=0$ and 
$y=\pi R/2$ compatible with the (different) $N=1$ supersymmetries 
at each of the fixed points. The FI terms in eq.~(\ref{eq:L-D}) 
can indeed be there, as can be supersymmetric kinetic terms for 
the different fields or any other $N=1$ supersymmetric operator of 
higher dimension. If not inserted from the start, one has to expect 
them from the loop expansion. In turn, their effect on the 
calculation of the Higgs potential has to be discussed along the 
lines of ref.~\cite{Barbieri:2001vh}.

The model, being based on a non-renormalizable Lagrangian, is 
defined in terms of a cutoff scale $M$, which we take to be the scale at 
which the perturbative expansion in the top Yukawa coupling ceases to 
make sense. It is $M \simeq 5/R$, as can be seen from an actual 
perturbative calculation or from naive dimensional analysis. We 
assume that perturbativity is maintained up to $M$, even after the 
inclusion of all possible other terms in the Lagrangian. This  
limits the effects of the various counterterms mentioned above on the 
Higgs potential. The closeness of $M$ to $1/R$ should not be viewed as 
an obstacle. In the chiral Lagrangian the relation of $m_{\rho}$ to 
$f_{\pi}$ is not very different and yet the usefulness of the 
chiral Lagrangian itself is not disputable.

Suppose that we use
\begin{equation}
  \Lambda \simeq M \simeq \frac{5}{R},
\end{equation}
to estimate the size of the $\xi$ term in eq.~(\ref{eq:xi}). 
We get $\xi \simeq 0.03/R^2$, which gives effects well within the 
uncertainties already discussed in ref.~\cite{Barbieri:2001vh} both 
on the Higgs potential and, a fortiori, on the superpartner spectrum. 
Radiatively generated brane kinetic terms, and higher loop corrections, 
give larger effects, as quoted above.

\begin{figure}[t]
\centerline{\epsfxsize=9cm \epsfbox{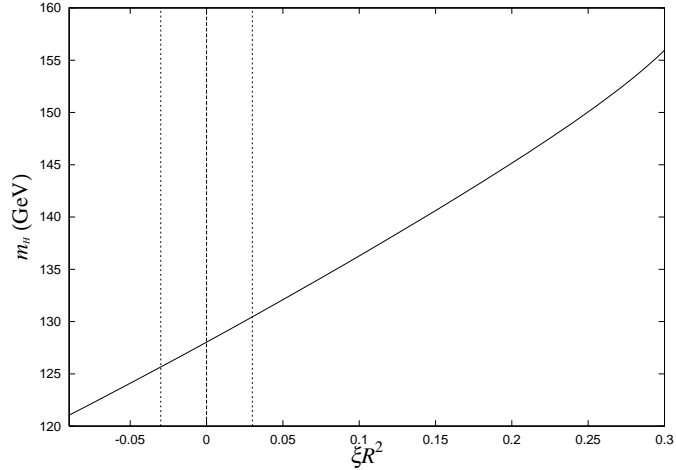}}
\caption{The physical Higgs boson mass $m_H$ as a function of $\xi R^2$.} 
\label{fig:plot-mh}
\end{figure}

\begin{figure}[t]
\centerline{\epsfxsize=9cm \epsfbox{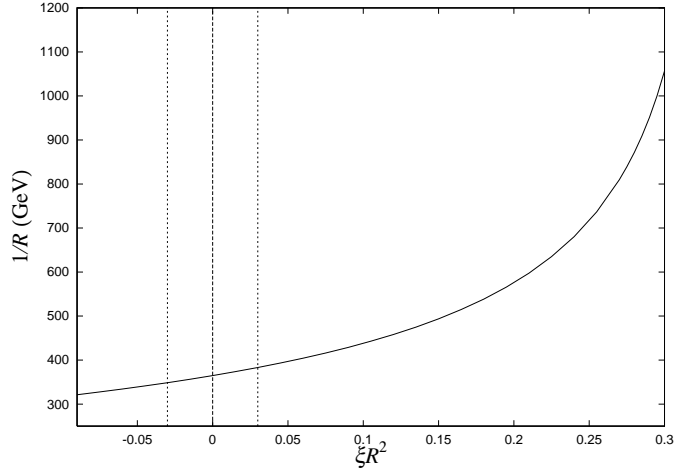}}
\caption{The compactification scale $1/R$ as a function of $\xi R^2$.} 
\label{fig:plot-Mc}
\end{figure}

The symmetries of the theory allow a tree level $D$-term as in 
eq.~(\ref{eq:L-D}), although with differing magnitudes on the two branes. 
This introduces a correction to the Higgs squared mass parameter
\begin{equation}
  \delta m_{\phi_H}^2 (\xi) = {g' \over 2} \xi, 
\end{equation}
where $\xi$ is now the average value of the $D$-term on the two branes. 
The sign of such a term is unknown.  It should be compared with the 
finite top loop contribution to the Higgs potential, which gives 
a mass squared
\begin{equation}
  \delta m_{\phi_H}^2 ({\rm top}) 
    = - \frac{63 \zeta(3)}{8 \pi^4} \frac{y_t^2}{R^2}
    \simeq -\frac{0.08}{R^2}.
\end{equation}
If $\delta m_{\phi_H}^2 (\xi)$ were positive and bigger than 
$|\delta m_{\phi_H}^2 ({\rm top})|$ its presence would prevent symmetry 
breaking.  For other values of $\xi$, we have minimized the Higgs potential
and predict the Higgs mass and $1/R$ as a function of the dimensionless 
parameter $\xi R^2$, as shown in figs.~\ref{fig:plot-mh} and 
\ref{fig:plot-Mc}. Vertical dashed lines at $\xi R^2 = \pm 0.03$ show the 
effects to be expected from the radiative FI term. The experimental limit 
on the top squark mass requires $\xi R^2$ to be larger than about $-0.1$, 
and electroweak symmetry is broken only if $\xi R^2$ is less than 
about 0.5.  The figures are not extended to values of $\xi R^2$ above 
0.3 because corrections to the top KK masses from $\xi$ have not been 
included in the calculation of the Higgs potential, and in this region 
they become important.
We note that a partial cancellation can take place between 
$\delta m_{\phi_H}^2 ({\rm top})$ and $\delta m_{\phi_H}^2 (\xi)$, with a 
corresponding increase in $1/R$ and $m_H^2$ itself. As $\xi R^2$ approaches
the maximum value consistent with electroweak symmetry being broken, 
larger values of $1/R$ result, but at the price of an increasingly precise 
cancellation among the two contributions to the Higgs mass. Very large 
values of $1/R$ are therefore disfavored, although even in this case 
there is a strict upper limit on the Higgs mass in the region of 180 GeV.
The hypercharge gauge interaction is weakly coupled at the cutoff scale 
$M$, so that $\xi$ cannot be estimated using strong coupling arguments.
A totally naive guess, $\xi \simeq g' M^2$, would be clearly excluded. 
The theory above $M$ must lead to suppressed tree-level
values for certain brane interactions, including the FI term and 
flavor-dependent kinetic terms for the light generations.

A value of $\xi R^2 \simeq 0.3$ raises the compactification 
scale by about a factor of 3, thereby removing the fine tuning needed to 
satisfy the experimental constraint on the $\rho$ parameter found in 
ref.~\cite{Barbieri:2001vh}.  Such a value of $\xi R^2$ is an order of 
magnitude larger than the radiative contribution, but over an order of 
magnitude smaller than a naive guess in 5D, and does not require 
fine tuning in the Higgs mass squared parameter. 

For values of $\xi$ which are not too large we find analytic 
approximations:
\begin{equation}
  m_H^2(\xi) \simeq m_H^2(0) + 
    M_Z^2 \Bigl( \cos[\pi R(0) m_t] - \cos[\pi R(\xi) m_t] \Bigr)
    + g' \xi \Bigl( 1 - \cos[\pi R(\xi) m_t] \Bigr),
\end{equation}
and
\begin{equation}
  \frac{1}{R(\xi)} \simeq \frac{1}{R(0)}
    \left( 1 + \frac{g'\xi}{M_Z^2} \right)^{\frac{1}{4}}.
\end{equation}
Furthermore, for $\xi$ sufficiently small, 
as in the one loop calculation, the predictions
made in ref.~\cite{Barbieri:2001vh} still hold:
\begin{eqnarray}
  m_H   &=& 127 \pm 8~{\rm GeV}, \\
  1/R   &=& 370 \pm 70~{\rm GeV}.
\end{eqnarray}

The role of the FI term in the Higgs potential is to provide an 
additional contribution to the Higgs mass squared parameter. In 
the case that this contribution is positive, the effect is 
equivalent to a deformation in the translation orbifold boundary 
condition discussed in ref.~\cite{Barbieri:2001dm}. Indeed, one can 
ask what measurement will provide a distinction between these theories.
The answer lies in the details of the scalar superpartner spectrum 
with masses near $1/R$. The boundary condition deformation leads 
to a universal shift in the scalar masses, while the FI term 
leads to shifts that depend on the hypercharge of the scalar.

The $D_Y$ field has a kinetic coupling to the real scalar field 
$\sigma$ of the hypercharge chiral adjoint field: $D_Y \partial_y \sigma$. 
On performing a KK expansion one discovers that the effects of 
the FI term coupled to $D_n$, for $n \neq 0$, are canceled by vacuum 
expectation values (VEVs) of $\sigma_n$. The physical effects of 
the FI term discussed above all result from the coupling to the 
zero mode $D_0$. The VEV of $\sigma_n$ leads to mass mixing amongst 
KK modes from the gauge interaction $g' [X \Sigma X^c]_{\theta^2}$ 
for any hypermultiplet $(X,X^c)$. This leads to a violation of 
momentum in the fifth dimension allowing single production of 
excited KK modes, such as $g_0g_0 \rightarrow \bar{q}_0 q_2$, and 
new decay modes of the excitations, such as $q_2 \rightarrow Z_0 q_0$,
where subscripts label the KK modes.

In the constrained standard model introduced in ref.~\cite{Barbieri:2001vh},
electroweak symmetry is broken radiatively via a finite 1 loop contribution 
involving the top quark and its superpartners and KK resonances. 
Corrections to this picture arise from supersymmetric brane interactions. 
There is a quadratically divergence brane FI term, as pointed out in 
ref.~\cite{Ghilencea:2001bw}, but this leads to only a 2\% correction in
the Higgs mass, which is smaller than other corrections.  It is perhaps
surprising that a 1 loop quadratic divergence is so mild relative to
the finite 1 loop top quark contribution. This results from several
factors: the top Yukawa coupling is larger than the hypercharge coupling, 
there is a color factor of 3, the Yukawa couplings of the KK towers 
are $\sqrt{2}$ times larger than that of the zero mode, and finally 
the cutoff of the theory is only about a factor of 5 above the 
compactification scale.  A tree level brane FI term could be
present. Since the hypercharge coupling is highly perturbative even at
the cutoff, it seems likely to us that this tree contribution from
physics at the cutoff is comparable to the quadratically divergent
radiative correction and therefore also negligible. 
However, if it is larger by an order of magnitude it could lead to
significant changes in the predictions of the theory as shown in  
figs.~\ref{fig:plot-mh} and \ref{fig:plot-Mc}.

\vspace{0.5cm}

{\bf Acknowledgements}

Y.N. thanks the Miller Institute for Basic Research in Science 
for financial support.
This work was supported by the ESF under the RTN contract 
HPRN-CT-2000-00148, the Department of Energy under contract 
DE-AC03-76SF00098 and the National Science Foundation under 
contract PHY-95-14797.

\vspace{0.5cm}

{\bf Appendix}

In this appendix we comment on theoretical issues about the generation 
of the FI term. We have seen that brane-localized operators, 
eq.~(\ref{eq:L-D}), are radiatively generated in the model of 
ref.~\cite{Barbieri:2001vh}.  The coefficients of the two FI terms on 
$y=0$ and $\pi R/2$ branes are the same, so that in the 4D picture 
the Lagrangian is given by
\begin{equation}
  {\cal L}_{\rm 4D} = \sqrt{2}\, \xi 
  \left( \frac{1}{\sqrt{2}} D_0 + D_2 + D_4 + \cdots \right),
\label{eq:D-a-h}
\end{equation}
where $D_n$ is the $n$-th KK mode of $D_Y$ with ``mass'' $2 n/R$.  
The question is what happens for the FI term if we take the 
supersymmetric limit of the theory.  Does the FI term remain non-zero?

To answer this question, let us describe our model using 
$Z: y \rightarrow -y$ and $T: y \rightarrow y + \pi R$ rather than 
$Z: y \rightarrow -y$ and $Z': (y-\pi R/2) \rightarrow -(y-\pi R/2)$.  
Then, our model corresponds to taking $Z$ as the ``standard'' $Z_2$ 
parity reducing 5D $N=1$ supersymmetry to 4D $N=1$ supersymmetry, and 
$T$ as a direct product of the SS rotation with the SS parameter 
$\alpha = 1/2$ and the overall negative sign for the Higgs 
hypermultiplet.  In the notation of ref.~\cite{Barbieri:2001dm},
it is written as
\begin{eqnarray}
  Z &=& \Sigma_3 \otimes 1, \\
  T &=& e^{2\pi i \alpha \sigma_2} \otimes -1,
\end{eqnarray}
where $\alpha = 1/2$ corresponds to the model of ref.~\cite{Barbieri:2001vh}.

Suppose we take SS parameter $\alpha$ to be equal to zero.  
Then, 4D $N=1$ supersymmetry remains unbroken, and both Higgs 
and Higgsino KK towers have mass $(2 n+1)/R$.  In this case, one might 
conclude that no FI term is generated since the matter content is 
completely vector-like; we have full hypermultiplet states, 
$h, h^c, \tilde{h}, \tilde{h}^c$, at each KK level.  
However, the situation is not so simple.
Although the matter content is vector-like, the interactions are not; 
$D_n$ ($n$:odd) interactions do not have a charge conjugation symmetry.
As a consequence, non-vanishing brane-localized FI terms are generated 
even in this supersymmetric case, $\alpha = 0$.
Indeed, a simple calculation shows that the terms of the form 
\begin{equation}
  {\cal L}_{\rm eff} = 
    {\xi \over \sqrt{2}} \left( \delta(y)  - \delta(y-\pi R/2) \right) D_Y,
\end{equation}
are generated radiatively.  In the 4D picture, this is 
\begin{equation}
  {\cal L}_{\rm 4D} = \sqrt{2}\, \xi 
    \left( D_1 + D_3 + D_5 + \cdots \right).
\label{eq:D-a-0}
\end{equation}
It is important to realize that this special form of the FI terms is 
guaranteed by a symmetry; in the 5D picture there is a charge 
conjugation symmetry which is accompanied by a spacetime reflection 
with respect to $y = \pi R/4$, and it allows only brane-localized 
FI terms with opposite coefficients at $y=0$ and $\pi R/2$.  
Incidentally, if $\alpha$ takes some arbitrary values, 
the situation is between the two extreme cases $\alpha = 0$ and $1/2$; 
the size of the FI terms on the two branes are different in general.

Thus, we conclude that, if we have only one hypermultiplet in the bulk, 
brane-localized FI terms are always generated (even if supersymmetry 
is not broken). However, there needs to be a slight care for this statement.
In the supersymmetric case of $\alpha = 0$, the generated terms do not 
contain the FI term for the unbroken (zero-mode) $U(1)$ hypercharge, 
in contrast with the case of $\alpha \neq 0$ (see eqs.~(\ref{eq:D-a-h}, 
\ref{eq:D-a-0})).  This is a reasonable result since
the appearance of a FI term is deeply related 
to the $U(1)$-(grav.)$^2$ anomaly in the usual 4D supersymmetric theories.
In 4D theories with $U(1)$-gravitational anomaly canceled, 
the FI term is never generated unless we break either supersymmetry 
or the $U(1)$.  In the present case with $\alpha = 0$, the ``FI term'' 
appeared in the supersymmetric limit.  However, it is a 
brane-localized term and not a true FI term in the 5D sense.
In other words, in the 4D picture the generated FI terms are only 
for higher KK modes and not for the zero mode.  Since the $U(1)$ 
symmetries corresponding to the higher KK towers are non-linearly 
realized (spontaneously broken), the generation of these FI terms 
do not conflict with the above general theorem in the 4D supersymmetric 
theories, nor break supersymmetry since the FI terms for $D_n$ ($n>0$)
are completely absorbed by the expectation values for the physical 
scalar field $\sigma_n$ coming from the gauge multiplet in 5D.

\end{document}